\documentclass[showpacs,preprintnumbers,superscriptaddress,amsmath,amssymb]{revtex4}

\usepackage{graphicx}
\usepackage{dcolumn}
\usepackage{bm}

\usepackage{dsfont}
\usepackage{amsmath}
\usepackage{amssymb}
\usepackage{amsthm}
\usepackage{amscd}
\usepackage{hyperref}

\newcommand{\be}{\begin{equation}}
\newcommand{\ee}{\end{equation}}
\newcommand{\bea}{\begin{eqnarray}}
\newcommand{\eea}{\end{eqnarray}}

\newcommand{\integer}{\relax{\rm I\kern-.18em N}}

\newcommand{\addrMOSKAU}{Institute for Theoretical and Experimental Physics, B.
Cheremushkinskaya 25, 117259 Moscow, Russia}
\newcommand{\addrTUEBINGEN}{Institut f\"{u}r Theoretische Physik, T\"{u}bingen
Universit\"{a}t, Auf der Morgenstelle 14, D-72076 T\"{u}bingen, Germany}
\begin{document}

\title{Comment on "Dynamics of nuclear fluid. VIII. Time-dependent Hartree-Fock approximation from a classical point of view"}


\author{M. I. Krivoruchenko}
\affiliation{\addrMOSKAU}
\affiliation{\addrTUEBINGEN}

\author{B. V. Martemyanov}
\affiliation{\addrMOSKAU}
\affiliation{\addrTUEBINGEN}

\author{C. Fuchs}
\affiliation{\addrTUEBINGEN}

\begin{abstract}

The phase-space paths introduced by Cheuk-Yin Wong in Phys. Rev. {\bf C25}, 1460 (1982) and discussed recently in the literature can be used for
calculation of evolution of the Wigner function to first order in the time increments only.
The first-order solutions are helpful to determine the phase-space Green function in the
framework of the phase-space path integral method.

\end{abstract}


\maketitle

Quantum trajectories exist
in the de Broglie - Bohm theory \cite{DeBrog,DBOHM,DBB} and appear
in the framework of the deformation quantization
\cite{WEYL1,WEYL2,WIGNER,GROE,MOYAL,BARL} as the Weyl's symbols of the Heisenberg operators
of canonical coordinates and momenta \cite{OSBOR,MCQUA,KRFA,DIAS,KFF}. In the de
Broglie - Bohm theory, the particle trajectories play an important role in the interpretation
of measurements, whereas within the deformation quantization framework quantum trajectories 
have properties assigned to characteristic lines of first-order partial differential
equations (PDE). The evolution equation for the Wigner function is the infinite-order PDE, nevertheless,
it can be solved with the help of quantum characteristics. The interest to particle trajectories is motivated
by the fact that they remain striking but intuitive feature of transport models in heavy-ion collisions
\cite{AICHE,Lehmann:1995sz,Bass:1998ca,Fuchs:2005zg}.

Already some time ago, C. Y. Wong \cite{Won82} proposed a new kind of quantum
trajectories for solving the evolution problem of the Wigner function. Recently
these ideas were discussed in Refs. \cite{Mor00,WONG}.

We wish to point out that the phase-space trajectories of Ref. \cite{Won82} can be used to find the evolution of the Wigner function to the first order in time increments only.

Consider the double Fourier transform of the Wigner function $W({\mathbf q},{\mathbf p},t)$:
\begin{equation}
W({\mathbf q},{\mathbf p},\tau) = \int{
\frac{d{\mathbf s}d{\mathbf p}^{\prime}}{(2\pi \hbar)^3} \exp(\frac{i}{\hbar} {\mathbf s}({\mathbf p} - {\mathbf p}^{\prime}))
W^{\prime}({\mathbf q},{\mathbf p}^{\prime},\tau)
}.
\label{DFT}
\end{equation}
We restrict ourselves to Hamiltonian functions of the form
\begin{equation}
H({\mathbf q},{\mathbf p}) = \frac{1}{2}{\mathbf p}^2 + V({\mathbf q}).
\label{HAMI}
\end{equation}
Using representation (\ref{DFT}), the quantum Liouville equation
\begin{equation}
\frac{\partial }{\partial \tau}W({\mathbf q},{\mathbf p},\tau) = -
W({\mathbf q},{\mathbf p},\tau) \wedge H({\mathbf q},{\mathbf p})
\label{QLE}
\end{equation}
where $\wedge$ is the Moyal bracket \cite{GROE,MOYAL,BARL} can be represented in the form
\begin{equation}
0 = \int
\frac{d{\mathbf s}d{\mathbf p}^{\prime}}{(2\pi \hbar)^3} \exp(\frac{i}{\hbar} {\mathbf s}({\mathbf p} - {\mathbf p}^{\prime}))
(\frac{\partial }{\partial \tau}  + {\mathbf p} \frac{\partial }{\partial {\mathbf q}}
- \frac{i}{\hbar}(
V({\mathbf q} + \frac{{\mathbf s}}{2}) - V({\mathbf q} - \frac{{\mathbf s}}{2})))W^{\prime}({\mathbf q},{\mathbf p}^{\prime},\tau).
\label{GRAN}
\end{equation}

It is tempting to require
\begin{equation}
0 \stackrel{?}= (\frac{\partial }{\partial \tau}  + {\mathbf p} \frac{\partial }{\partial {\mathbf q}} -
\frac{i}{\hbar}(V({\mathbf q} + \frac{{\mathbf s}}{2}) - V({\mathbf q} - \frac{{\mathbf s}}{2})))W^{\prime}({\mathbf q},{\mathbf p}^{\prime},\tau).
\label{GRAN-1}
\end{equation}
This is the first-order PDE. It can be solved using the method of characteristics:
\begin{equation}
W^{\prime}({\mathbf q},{\mathbf p}^{\prime},\tau) = W^{\prime}({\mathbf q} - {\mathbf p}\tau,{\mathbf p}^{\prime},0)
\exp({\frac{i}{\hbar}\int_{0}^{\tau}d\tau^{\prime}(V({\mathbf q} - {\mathbf p}(\tau - \tau^{\prime}) + \frac{{\mathbf s}}{2}) - V({\mathbf q} - {\mathbf p}(\tau - \tau^{\prime}) - \frac{{\mathbf s}}{2})))}).
\label{GRAN-2}
\end{equation}
We observe that Eq.(\ref{GRAN-1}) and its solution Eq.(\ref{GRAN-2})
depend on ${\mathbf s}$ and ${\mathbf p}$.

Given the dependence on ${\mathbf s}$ enters into $W^{\prime}({\mathbf q},{\mathbf p}^{\prime},\tau)$,
equation $W({\mathbf q},{\mathbf p},\tau) \equiv W^{\prime}({\mathbf q},{\mathbf p},\tau)$
does not hold anymore, however, Eq.(\ref{GRAN}) is still valid.
The appearance of ${\mathbf s}$ in $W^{\prime}({\mathbf q},{\mathbf p}^{\prime},\tau)$ does not violate the equivalence of Eqs.(\ref{QLE}) and (\ref{GRAN}).
However, given the dependence on ${\mathbf p}$ enters $W^{\prime}({\mathbf q},{\mathbf p}^{\prime},\tau)$, Eq.(\ref{GRAN}) becomes distinct from Eq.(\ref{QLE}). 


There is therefore the obvious shift in the definition of $W^{\prime}$ between Eqs.(1) - (5), since $W^{\prime}$ depends 
in (5) on a parameter. This parameter, $\textbf{p}$, is not strictly related to the same notation in Eq.(1) on which 
$W$ depends but $W^{\prime}$ does not. The arguments leading to Eqs.(\ref{GRAN-1}) and (\ref{GRAN-2}) are thereby not founded.

On the other hand, suppose that Eq.(\ref{GRAN-2}) holds nevertheless. Consider the evolution of
one-dimensional oscillator with potential $V(q) = \frac{1}{2}q^2$.
The Wigner function of the ground state has the form
\begin{equation}
W(q,p) = 2\exp(-q^2/\hbar - p^2/\hbar).
\label{WINGE}
\end{equation}
Applying Eqs. (\ref{DFT}) and (\ref{GRAN-2}), we obtain
\begin{equation}
W(q,p,\tau) \stackrel{??}= 2\exp(-(q + p\tau)^2/\hbar - (p - q\tau - p\tau^2/2)^2/\hbar).
\label{TDWF}
\end{equation}
The Wigner function of stationary states does not depend on time. This particular contradiction
proves that Eqs.(\ref{GRAN-1}) and (\ref{GRAN-2}) are incorrect for finite $\tau$.

The expression (\ref{TDWF}) holds however to order $O(\tau)$. One can verify, indeed,
that Eq.(\ref{GRAN-2}) upon the double Fourier transform (\ref{DFT}) gives the Wigner function correct
to order $O(\tau)$.

In Ref. \cite{WONG} a numerical example is considered to compare the exact evolution of the Wigner function
with the Wigner function calculated by the infinitesimal step-by-step evolution according to Eq.(\ref{GRAN-2}).
A discrete form of the phase-space path integral \cite{BLEAF1,MARI} is calculated \textit{per se}.
Equation (\ref{GRAN-2}) in its infinitesimal form leads to the representation of the path integral
discussed by Leaf \cite{BLEAF1} and Marinov \cite{MARI}. The first-order solution is therefore useful to obtain
the correct numerical results \cite{WONG}.

The role of trajectories entering the phase-space path integral
is distinct from the role of characteristics in phase space. Particle trajectories of the
de Broglie - Bohm theory depend on quantum states. In order
to find them one has to know the wave function.
The phase-space trajectories appearing as the Weyl's symbols of the Heisenberg operators
of canonical coordinates and momenta are currently the only known objects which can be treated
as quantum characteristics. They comprise the complete information on the evolution of quantum systems and can be used to find the evolution of the Wigner function for finite time intervals.

\vspace{1mm}
The authors wish to acknowledge useful discussions with Prof. Amand Faessler. 


\end{document}